\documentclass[aps,prl,showpacs,preprintnumbers,twocolumn]{revtex4}
\usepackage{bm}
\usepackage{latexsym}
\usepackage{dcolumn}
\usepackage{amsfonts,amssymb}
\usepackage{graphicx,epsfig}
\usepackage{psfrag}
\usepackage{amsmath,amssymb}

\def\be {\begin{equation}}
\def\ee {\end{equation}}
\def\ba {\begin{eqnarray}}
\def\ea {\end{eqnarray}}
\def\nn {\nonumber}
\def\a  {\alpha}

\def\d  {\delta}

\def\k  {\kappa}

\def\O  {\Omega}
\def\p  {\pi}

\def\t  {\tau}
\def\la {\label}
\def\le {\left}
\def\ri {\right}
\def\pa {\partial}
\def\f {\frac}

\def\no {\noindent}
\def\bi {\begin{itemize}}
\def\ei {\end{itemize}}

\def\laq{\hbox{~}\raise 0.4ex\hbox{$<$}\kern -0.8em\lower 0.62ex\hbox{$\sim$}\hbox{~}}
\def\gaq{\hbox{~}\raise 0.4ex\hbox{$>$}\kern -0.7em\lower
  0.62ex\hbox{$\sim$}\hbox{~}}

\def\beq{\begin{equation}}
\def\eeq{\end{equation}}
\def\br{\begin{eqnarray}}
\def\er{\end{eqnarray}}
\def\benu{\begin{enumerate}}
\def\eenu{\end{enumerate}}
\def\nn{\nonumber} 
\def\pa{{\partial}}
\def\l{\left}
\def\r{\right}    

\begin{document}
\title{Where are the black hole entropy degrees of freedom ?}

\author{Saurya Das$^{\S}$ and S. Shankaranarayanan$^{\P}$}
\affiliation{$^{\S}$ Dept. of Physics, University of Lethbridge,
4401 University Drive, Lethbridge, Alberta T1K 3M4, Canada}
\email{saurya.das@uleth.ca}
\affiliation{$^{\P}$
Max-Planck-Institut f\"ur Gravitationphysik, 
Am M\"uhlenberg 1, D-14476, Potsdam, Germany
}
\email{shanki@aei.mpg.de}

\begin{abstract}
Understanding the area-proportionality of black hole entropy (the
`Area Law') from an underlying fundamental theory has been one of the
goals of all models of quantum gravity. A key question that one asks
is: {\it where are the degrees of freedom giving rise to black hole
entropy located?} Taking the point of view that entanglement between
field degrees of freedom inside and outside the horizon can be a
source of this entropy, we show that when the field is in its ground
state, the degrees of freedom near the horizon contribute most to the
entropy, and the area law is obeyed. However, when it is in an excited
state, degrees of freedom far from the horizon contribute more
significantly, and deviations from the area law are observed. In other
words, we demonstrate that horizon degrees of freedom are responsible
for the area law.
\end{abstract}

\maketitle

The area proportionality of black hole entropy (the `{\it Area Law}' (AL)),
\br
S_{_{\rm BH}} = \frac{A_H}{4 \ell_P^2 } ,\qquad
\le(\ell_{_P} \equiv \sqrt{{\hbar G}/{c^3}} =  \mbox{Planck length} \ri)
\la{al1}
\er
which differs from the volume proportionality of familiar
thermodynamic systems, has been conjectured to be more fundamental in
some sense (the {\it Holographic Hypothesis}). Black holes are also
regarded as theoretical laboratories for quantum gravity.  Thus,
candidate models of quantum gravity, such as string theory and loop
quantum gravity, have attempted to derive the `macroscopic' AL
(\ref{al1}) by counting `microscopic' degrees of freedom (DOF), using
the Von-Neumann/Boltzmann formula \cite{stringsetc}:
\br
S = - Tr[\rho \ln(\rho)]~= \ln \Omega, \qquad \quad k_B =1
\label{vonneumann}
\er
where $\rho$ and $\Omega$ correspond to the density matrix and number of
accessible micro-states, respectively.  Depending on the approach, one
either counts certain DOF on the horizon, or abstract DOF related to
the black hole, and there does not appear to be a consensus about
which DOF are relevant or about their precise location \cite{wald}.

In this article, we attempt to answer these questions in a more
general setting, which in fact, may be relevant in any approach.  We
adopt the point of view that the entanglement between quantum field
DOF lying inside and outside of the horizon leads to black-hole
entropy. It was shown in Refs. \cite{bkls,sred} (see also,
Refs. \cite{eisert,arom}) that the entanglement entropy of a massless
scalar field propagating in flat space-time (by tracing over a
spherical region of radius $R$) is proportional to the area of the
sphere
\beq
S_{_{\rm ent}} = 0.3~\le(\frac{R}{a}\ri)^2 \qquad a \mbox{~is the UV cutoff} 
\,. 
\la{gslaw}
\eeq
This suggests that the area law is a generic feature of entanglement,
and acquires a physical meaning in the case of black-holes, the
latter's horizon providing a natural `boundary' of the region to trace
over.  Note that Eq.(\ref{gslaw}) will continue to hold if the region
outside the sphere is traced over instead.

The relevance of $S_{_{\rm ent}}$ to $S_{_{\rm BH}}$ can also be
understood from the fact that both entropies are associated with the
existence of the {\it horizons} \cite{Jacobson:2005}. Consider a scalar
field on a background of a collapsing star.  At early times, there is
no horizon, and both the entropies are zero. However, once the horizon
forms, $S_{_{\rm BH}}$ is non-zero, and if the scalar field DOF inside
the horizon are traced over, $S_{_{\rm ent}}$ obtained from the
reduced density matrix is non-zero as well.

In Refs. \cite{bkls,sred}, along with the fact that calculations were
done in flat space-time, a crucial assumption was made that the scalar
field is in the {\it Ground State} (GS). In Refs. \cite{ads,sdssprd},
the current authors studied the robustness of the AL by relaxing the
second assumption, and showed that for {\it Generic
Coherent States} (GCS) and a class of {\it Squeezed States} (SS), the
law continues to hold, whereas for the {\it Excited States} (ES), one obtains:
{\small
\beq
S_{_{\rm ent}} = \kappa \, \l(\frac{R}{a}\r)^{2 \alpha}~, 
\qquad \kappa={\cal O}(1)
\la{eslaw}
\eeq 
}
\hspace*{-0.25cm} where $\alpha<1$, and higher the excitation, the
smaller its value.  In this article, we attempt to provide a physical
understanding of this deviation from the AL, by showing that for ES,
DOF far from the horizon contribute more significantly than for the
GS. We also extend our flat space-time analyses for any $(3 + 1)-$D
spherically symmetric non-degenerate black-hole space-times
\cite{mukoh}.

We begin by considering a scalar field $\phi(x)$ propagating in a
Schwarzschild space-time
\footnote{The
perturbations of a $(3 + 1)$-dimensional static spherically symmetric
black holes result in two kinds -- axial and polar -- of gravitational
perturbations. The equation of motion of the axial perturbations is
same as that of a test scalar field propagating in the black-hole
background \cite{chandra,DS-new}. Hence, by computing $S_{_{\rm ent}}$ of the
test scalar fields, we obtain entropy of black-hole metric
perturbations.}:
{\small
\be
ds^2 = -f(r) dt^2 + \f{dr}{f(r)} + r^2 d\O_{_{2}}^2
\quad,\quad f(r)= 1 - \f{r_0}{r}~.
\la{sch1}
\ee
}
\hspace*{-0.25cm} where $r_0$ is the horizon.  Transforming to
Lema\^itre coordinates $(\tau,R)$ \cite{landau2}:
\ba
\la{Lemtrans}
\tau &=& t + r_0 \le[ \ln\le( \f{1-\sqrt{r/r_0} }{1+\sqrt{r/r_0}} \ri) 
+ 2\sqrt{\f{r}{r_0}}  \ri]  \\
R &=& \tau + \f{2~r^{\f{3}{2}}}{3\sqrt{r_0}} 
\quad \Longrightarrow \quad 
\frac{r}{r_0} = \le[\frac{3}{2}\frac{(R-\t)}{r_0}\ri]^{{2}/{3}} \, , \nn
\ea
%
%
the line-element (\ref{sch1}) becomes: 
{\small
\be 
ds^2= -d\t^2 + \le[ \f{3}{2} \frac{(R-\t)}{r_0}\ri]^{-\f{2}{3}}\!\!\!{dR^2} 
+ \le[\f{3}{2} \frac{(R-\t)}{r_0}\ri]^{\f{4}{3}}r_0^{\f{2}{3} } d\O^2 \, . 
\la{sch2}
\ee
}
Note that $R (\tau)$ is everywhere space-like (time-like), the metric
is non-singular at $r=r_0$ (corresponding to $3(R-\tau)/2r_0 = 1$) and 
the metric is explicitly time-dependent.
The Hamiltonian of a massless, minimally coupled scalar field in the
background (\ref{sch2}) is given by 
%
\br
\la{schham}  
H(\t)&=& \sum_{lm} \f{1}{2} \int_\tau^\infty \!\!\!\! dR 
\le[ \f{2 P_{lm}^{2}(\t,R)}{3(R-\t)} + \f{3 r}{2}(R-\t) \r . \\ 
& \times& \l.  
\le[\pa_R \phi_{lm}(\t,R)\ri]^2 
+ \sqrt[3]{\frac{r_0}{r}} \, \ell (\ell + 1) \phi_{lm}^2(\t,R) \r] \, , 
\nn
\er
where $P_{lm}(\tau,R)$ is the canonical conjugate momenta of the
spherically reduced scalar field $\phi_{lm}(\tau,R)$, such that
$[\hat{\phi}_{lm}(R,\t_0),\hat{P}_{lm}(R',\t_0)] = i \delta(R - R')$
and $\ell$ denotes angular momenta. Note that, in these coordinates,
the scalar field and its Hamiltonian explicitly depend on time.

Next, choosing a {\it fixed} Lema\^itre time (say, $\t_0 = 0$) and
performing the following canonical transformation
\cite{mw}
\be
P_{lm}(r) = \sqrt{r} \, \p_{lm}(r) \quad, \quad 
\phi_{lm}(r) = \f{\varphi_{lm}(r)}{r} \, ,
\ee
the Hamiltonian (\ref{schham}) transforms to:
{\small
\be
H(0) = \sum_{lm}\int_0^\infty \! \frac{dr}{2} 
\le[\p_{lm}^2(r) + r^2 \le[\pa_r \f{\varphi_{lm}}{r}\ri]^2
+ \frac{\ell (\ell + 1)}{r^2} \varphi_{lm}^2(r) \ri] \, 
\label{eq:FST-Hamil}
\ee
}
\hspace*{-0.25cm} which is simply the Hamiltonian of a free scalar
field in flat space-time!  Eq.(\ref{eq:FST-Hamil}) holds for {\it any}
fixed $\tau$, for which the results of Refs. \cite{sred,sdssprd},
namely relations (\ref{gslaw}) and (\ref{eslaw}), go through, provided
one traces over either the region $R \in [0, \f{2}{3} r_0)$ or
the region $R \in [\f{2}{3} r_0, \infty) $ \cite{mukoh}.
%
Extension to any non-degenerate spherically symmetric space-times is 
straightforward
\footnote{Ideally, one would like to fix the vacuum state at some
Lemaitre time $\t = 0$ and study the evolution of the modes at late
time leading to Hawking effect. Such an analysis is nontrivial for the
$(3 + 1)-D$ black-holes. Even in the case of $(1 + 1)-D$ black-holes,
this issue has been addressed with limited success \cite{Jacobson-2k}.
In the rest of the article, we will not consider these effects.}.


Remaining steps in the computation of $S_{_{\rm ent}}$ are:
\\ 1. {\it Discretize the Hamiltonian (\ref{eq:FST-Hamil})}, i. e.,
\br
H &=& \sum_{lm} \f{1}{2a} \sum_{j=1}^N
\le[ \p_{lm,j}^2 + \le(j+\f{1}{2}\ri)^2
\le( \f{\varphi_{lm,j}}{j}
- \f{\varphi_{lm,j+1}}{j+1}
\ri)^2 \r. \nn \\
& & \qquad \qquad \qquad 
+ \l. \f{l(l+1)}{j^2}~\varphi_{lm,j}^2 \ri]  \quad , \quad 
\label{disc1}
\er
where $\pi_{lm,j}$ denotes the conjugate momenta of $\varphi_{lm,j}$,
$L=(N+1)a$ is the box size and $a$ is the radial lattice size.  This
is of the form of the Hamiltonian of $N$ coupled HOs
\be
H = \f{1}{2} \sum_{i=1}^N p_i^2 
+\f{1}{2} \sum_{i,j=1}^N x_i K_{ij} x_j 
\qquad i,j=1,\dots,N
\la{coupledham1}
\eeq
with the interaction matrix elements $K_{ij}$ given by:
\ba
&& K_{ij} = 
\frac{1}{i^2} \le[l(l+1)~\delta_{ij} + \f{9}{4}~\d_{i1} \d_{j1} 
+ \le( N - \f{1}{2}\ri)^2 \d_{iN} \d_{jN}  \ri. \nn \\
&&\le.  + \le( \le(i+\f{1}{2}\ri)^2 + \le(i - \f{1}{2} \ri)^2
\ri) \d_{i,j(i\neq 1,N)} \ri]  \nn \\
&&-\le[ \f{(j+\f{1}{2} )^2}{j(j+1)} \ri] \delta_{i,j+1} -
\le[ \f{(i+\f{1}{2} )^2}{i(i+1)} \ri] \delta_{i,j-1} \, , 
\la{kij}
\ea
where the last two terms are the nearest-neighbor interactions.

\no 2. {\it Choose the state of the quantum field:} For GS
($r=1,\alpha_i=0$), GCS ($r=1,\a_i=$~arbitrary), or a class of SS
($\a=0,r=$~arbitrary), the N-particle wave function $\psi(x_1 \dots
x_N)$ is given by \cite{sdssprd}
\ba 
\psi(x_1 \cdots x_N) = \le|\f{\O}{\p^N} \ri| \exp\le[- \frac{1}{2}
\sum_i r \k_{Di}^{1/2} \le( {\underbar x}_i - \a_i \ri)^2\ri]\!.
\la{cswf2b} 
\ea
For ES (or 1-Particle state), the N-particle wave function
$\psi^{}(x_1 \dots x_N)$ is given by 
{\small
\ba
\psi^{}(x_1 \cdots x_N) = \le| \f{2 \O}{\p^N}\ri|^{\f{1}{4}}
\sum_{i=1}^N a_i k_{D i}^{\f{1}{4}} \, {\underbar x}_i 
\exp\le[ -\f{1}{2} \sum_{j} k_{D j}^{\f{1}{2}}~{\underbar x}_j^2 \ri] 
\ea
}
\noindent \hspace*{-8pt} where $a^T = \le( a_1,\dots, a_N \ri)$ are
the expansion coefficients (normalization requires $a^T a=1$). For our
computations, we choose $a^{T} = 1/\sqrt{o}(0, \cdots 0, 1 \cdots 1)$
with the last $o$ columns being non-zero. For details, see
Ref. \cite{sdssprd}.

\no
3. {\it Obtain the density matrix:} For an arbitrary wavefunction
$\psi(x_1,\dots,x_N)$, the density matrix --- tracing over first $n$
of the $N$ field points --- is given by:
\ba
\rho \le( t; t'\ri) &=& 
\int \prod_{i=1}^n~dx_i~\psi (x_1,\dots,x_n;t_1,\dots,t_{N-n}) \nn \\
&& \quad \times \quad \psi^\star (x_1,\dots,x_n;t_1',\dots,t_{N-n}') 
\la{denmatgen1}
\ea
where: $t_j \equiv x_{n+j}, t \equiv t_1,\dots,t_{N-n}, ~j=1..(N-n),
x^T = (x_1,\dots,x_n;t_1,\dots,t_{N-1}) = (x_1,\dots,x_n;t) $. This
step, in general, cannot be evaluated analytically and requires
numerical techniques. For GS/CS/SS, substituting (\ref{cswf2b}) in the
above expression and using the relation (\ref{vonneumann}) gives
Eq.~(\ref{gslaw}). For ES, on the other hand, this leads to the
relation (\ref{eslaw}) \cite{sdssprd}.

\begin{figure}[!htb]
\begin{center}
\epsfxsize 3.00 in
\epsfysize 3.00 in
\epsfbox{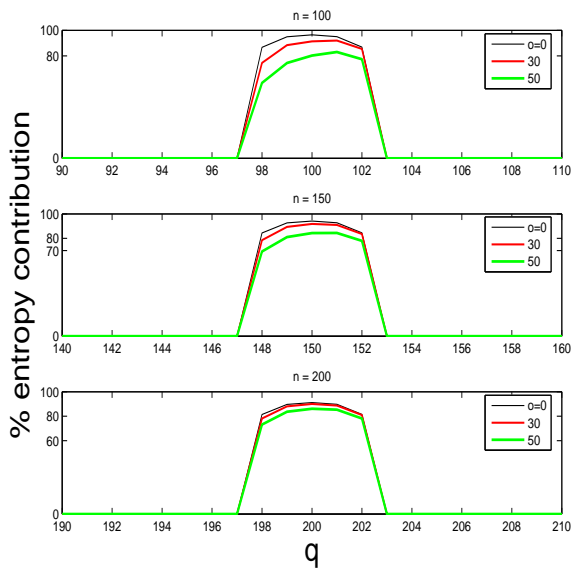}
\caption{Plot of the percentage contribution to the entropy for the GS
and ES as we move the window $q - 2 \leq i < q + 2$ from $q=2$ to
$q=n$, for $N=300$ and $n = 100, 150$ and $200$. The solid thin curve
is for GS $o = 0$ and bold (light) thick curve for $o = 30 (50)$.}
\label{fig:1}
\end{center}
\vspace*{-0.55cm}
\end{figure}

\begin{figure}[!htb]
\begin{center}
\epsfxsize 3.00 in 
\epsfysize 3.00 in 
\epsfbox{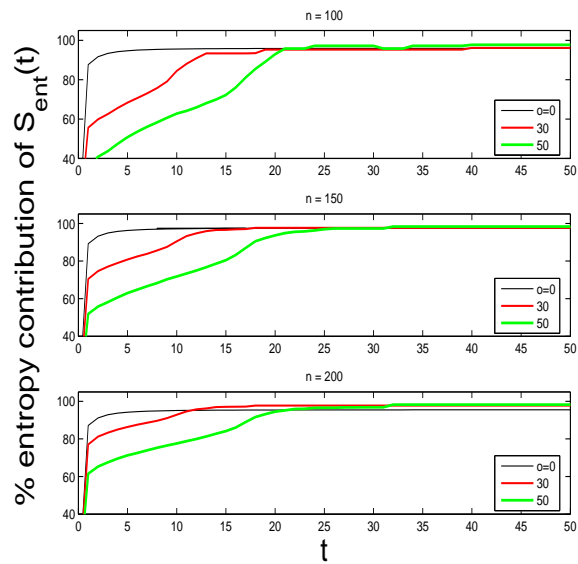}
\caption{Plot of the percentage contribution of $S_{ent}(t)$ for the
GS and ES. Here again, $N=300$, $n = 100, 150$ and $200$, and the
solid thin curve is for GS and bold (light) thick curve for $o = 30
(50)$.}
\label{fig:2}
\end{center}
\vspace*{-0.55cm}
\end{figure}

Now, to locate those DOF which are responsible for entropy, we take a
closer look at the interaction matrix (\ref{kij}). As mentioned
earlier, the last two terms are the nearest-neighbor interactions
between the oscillators and constitute entanglement. As expected, if
these terms are set to zero (by hand), $S_{\rm ent}$
vanishes. Instead, let us do the following: \\
\no (i) Set the interactions to zero (by hand) everywhere except in
the `window' ($ q-s \leq i \leq q + s$), with $ s \leq q \leq n-s$,
i.e., restrict the thickness of the interaction region to $2s+1$
radial lattice points, while moving it rigidly across from the origin
to horizon. In Fig. (\ref{fig:1}), we have plotted the percentage
contribution to the total entropy as a function of $q$ for the
GS/GCS/SS ($o = 0$, solid thin curve) and the ES ($o = 30 (50)$, bold
(light) thick curve). [We choose $N = 300$ and $n = 100, 150,
200$. The numerical error in the evaluation of the entropy is less
than $0.1 \%$.]
Fig. (\ref{fig:1}) shows that:
\begin{itemize}
\item When the interaction region does not include the horizon, the
entanglement entropy is zero and it is maximum if the horizon lies
exactly in its middle.  The first observation confirms that
entanglement between DOF inside and outside the horizon contributes to
entropy, while the second suggests that DOF close to the horizon
contribute more to the entropy compared to those far from it.
\item When the number of excited states is increased (i. e. $o = 30,
50$), the percentage contribution to the total entropy close to the
horizon is less compared to that of GS and the (bold/light thick)
curves are flatter.  These indicate that, for ES, there is a
significant contribution from the DOF far-away from the horizon.
\end{itemize}
\noindent 
(ii) To further investigate, we now set the interactions to zero (by
hand) everywhere except in the window $ p \leq i \leq n$, with the
horizon as its outer boundary, and vary the width of the window
$t\equiv n-p$ from $0$ to $n$.  For $t=n$, the total entropy is
recovered, while for $t=0$, i.e. zero interaction region, entropy
vanishes. In Fig. \ref{fig:2}, we have plotted the normalized
GS/GCS/SS and ES entropies [$S_{ent}(t)$] vs $t$ for $n = 100, 150$
and $200$. [Here again, the solid thin curve is for GS and bold
(light) thick curve for $o = 30 (50)$.] We infer the following:
\begin{itemize}
\item For the GS/GCS/SS, the entropy reaches the GS entropy for as
little as $t = 3$.  In other words, the interaction region
encompassing DOF close to the horizon contribute to most of the
entropy for the GS/GCS/SS.
\item In the case of ES, the entropy reaches the ES entropy only for
$t = 15 (20)$ [for $o = 30 (50)$]. This indicates that: (a) The DOF
far-away from the horizon have a greater contribution than that of GS
and (b) As the number of excited states increases, contribution
far-away from the horizon also increases.

This leads us to {\it one of the main conclusions} of this article:
the greater the deviation from the AL, the larger is the contribution
of the DOF far-away from the horizon.  It can be shown that the
density matrix for the ES is {\it more spread out} than for the GS.
\end{itemize}
%

\begin{figure}[!htb]
\begin{center}
\epsfxsize 3.00 in
\epsfysize 3.00 in
\epsfbox{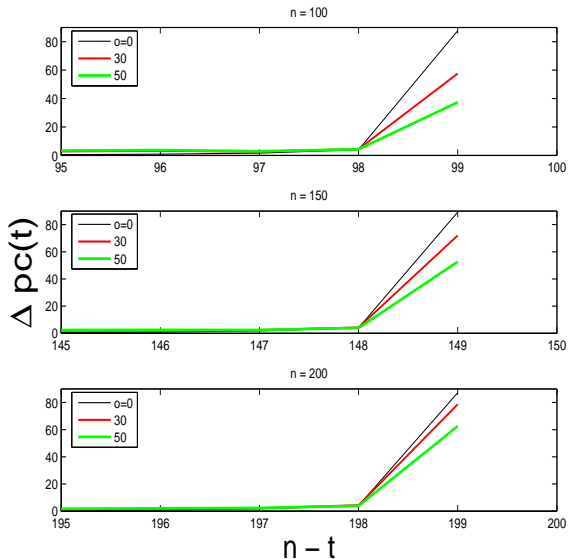}
\caption{Plot of $\Delta pc(t)$ vs $n- t$ for GS and ES. Here again,
$N=300$, $n = 100, 150$ and $200$, and the solid thin curve is for GS
and bold (light) thick curve for $o = 30 (50)$.}
\label{fig:3}
\end{center}
\vspace*{-0.55cm}
\end{figure}

\noindent (iii) To understand this further, let us define
{\small
$$\Delta pc(t) = pc(t) -
pc(t-1) \quad \mbox{where} \quad 
pc(t) = \f{S_{\rm ent}(t)}{S_{\rm ent}}\times 100 \, ,
$$
}
\hspace*{-0.25cm} where $pc(t)$ is the percentage contribution to the
total entropy by an interaction region of thickness $t$ and $\Delta
pc(t)$ is the percentage increase in entropy when the interaction
region is incremented by one radial lattice point. In other words,
$\Delta pc(t)$ is the slope of the curves in Fig. (\ref{fig:2}). In
Fig. (\ref{fig:3}), we have plotted $\Delta pc(t)$ vs $(n-t)$ for GS
and ES.  For the GS/GCS/SS, it is seen that when the first lattice
point just inside the horizon is included in the interaction region,
the entropy increases from $0$ to $85\%$ of the GS entropy.  Inclusion
of the next three points add another $9\%, 3\%, 1\%$ respectively. The
contributions to the entropy decrease rapidly and by the time the
$(n/3)^{th}$ point is included, entropy barely increases by a
hundredth of a percent!  For ES however, inclusion of one lattice
point adds $70(50)\%$, for $o=30(50)$, to the entropy, while the next
few points add $9\%,4\%,3\% \cdots$ respectively.  The corresponding slopes
are smaller. 

Figures (\ref{fig:2}) and (\ref{fig:3}) suggest the 
next key result of this article - most of the
entropy comes from the DOF close to the horizon. However, the DOF deep
inside must also be taken into account for the AL (\ref{gslaw}) to
emerge for the GS/GCS/SS, and the law (\ref{eslaw}) to emerge for the
ES.  These DOF affect the horizon DOF via the nearest-neighbor
interactions in (\ref{kij}). 

Our work clearly demonstrates the close relationship between the AL
and the horizon DOF, and that when the latter become less important,
the entropy scales as a power of area less than unity. This can be
understood from the following heuristic picture: taking the point of
view that $S_{_{BH}}$ is proportional to ${\cal N}$, the number of DOF
{\it on the horizon}, and that for the GS/CS/SS there is one DOF per
Planck-area, such that ${\cal N} \propto A_H$, the AL follows (This is
known as the {\it it-from-bit picture} \cite{wheeler}).  For ES
however, since this number is seen to get diminished, it will be given
by another function ${\cal N}=f(A)<A$. Now, since current results are
expected to be valid when $A\gg1$ (in Planck units) (such that
backreaction effects on the background can be neglected), it is quite
plausible that $f(A) \sim A^\alpha, ~\alpha<1$, just as we
obtain. Note that the above reasoning continues to hold when the
outside of the horizon is traced over.  Another way of understanding
our result is as follows: all interactions being of the
nearest-neighbor type [cf. Eq.(\ref{kij})], the degrees of freedom
deep inside the horizon influence the ones near the horizon (and hence
contribute to the entropy) only indirectly, i.e. via the intermediate
degrees of freedom. Evidently, their influence diminishes with their
increasing distance from the horizon. When they are excited however,
they have more energy to spare, resulting in an increased overall
effect.
Further investigations with superpositions of GS and ES are
expected to shed more light on this \cite{DS-new}.

We would like to thank A. Dasgupta, A. Roy, S. Sengupta and S. Sur for
discussions.  This work was supported in part by the Natural Sciences
and Engineering Research Council of Canada and the Perimeter Institute
for Theoretical Physics.

\end{document}